\documentclass[preprint,pre,floatfix]{revtex4}

\usepackage{graphicx}
\usepackage{dcolumn}
\usepackage{bm}
\usepackage{amsmath}
\usepackage{graphics}

\begin{document}

\title{Cytoskeleton mediated effective elastic properties of model
red blood cell membranes}

\author{Rui Zhang}
\author{Frank L. H. Brown}
\affiliation{Department of Chemistry and Biochemistry and Department
of Physics, University of California, Santa Barbara, California}

\date{\today}

\begin{abstract}
The plasma membrane of human red blood cells consists of a lipid bilayer
attached to a regular network of underlying cytoskeletal polymers.  We
model this system at a dynamic coarse-grained level, treating the bilayer as an
elastic sheet and the cytoskeletal network as a series of phantom entropic springs.
In contrast to prior simulation efforts, we explicitly account for dynamics of the
cytoskeletal network, both via motion of the protein anchors that attach the cytoskeleton
to the bilayer and through breaking and reconnection of individual cytoskeletal
filaments.  Simulation results are explained in the context of a simple
mean-field percolation model and comparison is made to experimental
measurements of red blood cell fluctuation amplitudes.

\end{abstract}


\maketitle

\section{Introduction}\label{sec:introduction}

\begin{figure}
\begin{center}
\includegraphics[width=0.45\textwidth]{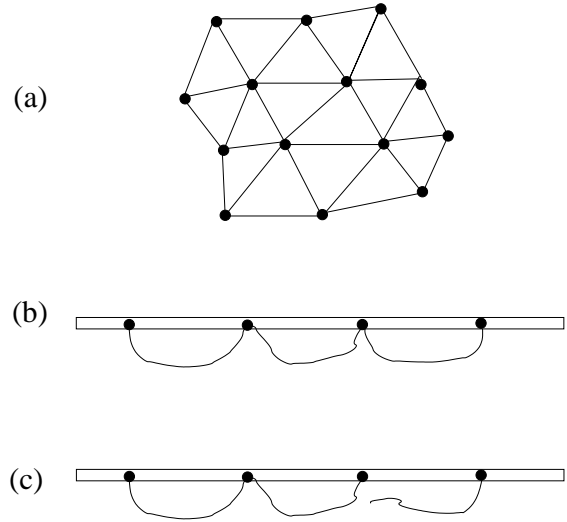}
\end{center}
\caption{A schematic rendering of the RBC membrane. (a): Top View:
Looking down on the cell with the lipid bilayer invisible for
clarity. The nodes of the cytoskeletal network are protein
complexes embedded in the lipid bilayer. The links are spectrin
tetramers, with approximate lateral end-to-end distance of $100
nm$ between the nodes. (b): Side View:   The protein anchors are
confined to the lipid bilayer, but may diffuse laterally. The
coils underneath the bilayer are spectrin tetramers.  The contour
length of the tetramers is approximately $200 nm$, so there is
considerable extra length coiled up below the membrane surface.
(c): One end of a link is disconnected from the node, possibly
with the help of ATP.} \label{fig:mesh}
\end{figure}

Membranes are essential components of all biological cells \cite{cell}.
In addition to their biological importance, lipid bilayers and
biomembranes have also attracted considerable attention from physicists
due to their fascinating and unusual properties \cite{Safran,Green_book}.
One particularly well studied system is the human red blood cell
(RBC) membrane.  The RBC membrane is a composite structure, consisting
of a lipid bilayer adhered to an underlying network of filamentous cytoskeletal
proteins (spectrin) via integral membrane protein anchors (see Fig.~\ref{fig:mesh}).
The spectrin network is observed to be quite regular \cite{EM}, with an approximate
hexagonal symmetry extending over the entire cell surface.  (It is worth
emphasizing that the spectrin based cytoskeletal network of RBCs is completely different
from the three dimensional actin networks common to other types
of animal cells \cite{cell}.)


Given the apparent simplicity of the RBC membrane, it is tempting
to attempt modeling with elementary elastic models.  Indeed, there
is a rich history of elastic RBC models to be found in the
literature
\cite{brochard,peterson,GovPRL,WortisPNAS,FournierPRL,RochalPRL}.
Without attempting a full historical review here, we comment that
no single elastic model has yet been identified that is capable of
reproducing the full set of experimental data available for RBC
membranes.  For example, while the work of Lim, Wortis and
Mukhopadhyay \cite{WortisPNAS} is capable of capturing the range
of observed RBC shapes (stomatocyte, discocyte, echinocyte and
non-main-sequence shapes as well) seen under various chemically
induced stresses (e.g. pH, salt, ATP, etc.), this model has not
been applied to explain the thermal fluctuation amplitudes
observed in RBC membranes.  And, while the models of Gov, Zilman
and Safran \cite{GovPRL} and Fournier, Lacoste and Raphael
\cite{FournierPRL} appear to do a good job fitting thermal
fluctuation data \cite{Zilker}, these models do not appear capable
of explaining mechanical deformation experiments on RBCs
\cite{Discher,Heinrich,SackmannNature}.  The failure of simple
models to consistently explain both thermal fluctuations and
mechanical deformation experiments has been recognized for years
\cite{peterson}.  One recent model does explain both types of data
within a single elastic model \cite{RochalPRL}, however this model
treats the spectrin network as an incompressible and homogeneous
viscoelastic plate coupled to the lipid bilayer.  It remains unclear
why such an approximation should suffice for the sparse cytoskeletal
network present in RBCs.  Additionally, this model seems too simplistic to
capture the full range of shape behaviors explained in reference
\cite{WortisPNAS}.

The models mentioned in the preceding paragraph make no mention of
the role of energy expenditure in the behavior of RBC membranes.
This despite the fact that it is known that RBC membranes possess
kinase and phosphatase activities capable of altering the properties of
spectrin and other network associated proteins
via (de)phosphorylation \cite{Birchmeier,Bennett}. And,
certain measurements of RBC fluctuation amplitudes show a
correlation between ATP concentration and fluctuation magnitude
\cite{Korenstein,Tuvia}.  One might argue that the difficulty in
fitting all RBC behavior to a single elastic model stems from the
fact that a truly comprehensive model must incorporate the effects
of energy expenditure by the cell in a realistic fashion.  Gov and
Safran \cite{nir_cyto} are the first to seriously consider active
energy expenditure within the RBC from a theoretical standpoint.
They have proposed that ATP induced phosphorylation and
dephosphorylation of the RBC cytoskeletal network leads to a continual dynamic
evolution of the integrity of the spectrin network. While this
picture remains hypothetical, without direct proof, it
is consistent with the general observations relating ATP
concentration to membrane fluctuation amplitudes.  Gov and Safran
(GS) \cite{nir_cyto} have used this picture to motivate a simple
picture for RBC fluctuations under the presence of ATP.  Local
breaking and reforming of the spectrin network is captured via
non-thermal forces imparted on an elastic membrane model. This
model has provided the first plausible explanation for the
viscosity dependence of RBC fluctuations \cite{Tuvia2}.

While elastic models  with proper accounting for non-thermal
energetics may eventually prove adequate in describing the long
wavelength physics of the RBC, it is clear that wavelengths near
or below the spectrin mesh size ($\sim 100 nm$) must be considered
within a more microscopic picture.  A recent model by Dubus and
Fournier (DF) \cite{Fournier} has extended the traditional elastic
modeling of RBC membranes to explicitly include the cytoskeletal
network at a molecular level of detail.  Within this model, the
spectrin network is considered as a completely regular hexagonal
network of phantom entropic springs attached to a fluid lipid
bilayer. Over wavelengths significantly longer that the spectrin
mesh size, the network so modeled becomes mathematically
equivalent to an imposed surface tension on the fluid bilayer.  At
wavelengths comparable to and shorter than network spacing, the
system behaves differently from a simple membrane with applied
tension.  This model was used to compute the spectrum of thermal
fluctuation amplitudes for the RBC membrane, but made no attempt
to account for non-thermal consumption of energy and only computed
thermal (non-dynamic) observables.

\begin{figure}
\begin{center}
\includegraphics[width=0.5\textwidth]{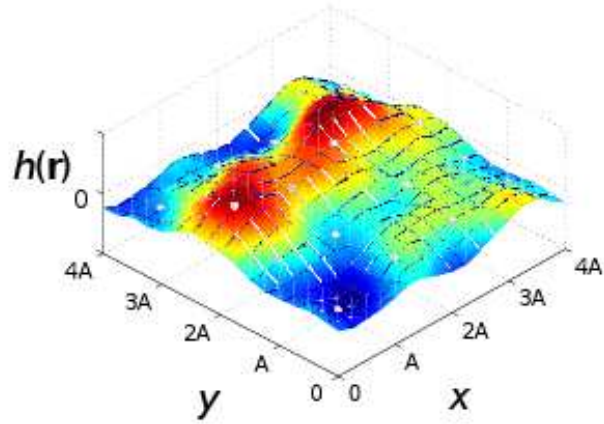}
\end{center}
\caption{(Color online). A snapshot the RBC membrane as treated in
our simulations. The lipid bilayer is connected to spectrin via a
series of laterally mobile anchoring proteins (white dots).  The
anchors are interconnected via harmonic potentials (not shown for
clarity), modeling the entire cytoskeletal network as a series of
interconnected entropic springs.  Our model does not account for
any steric repulsion between cytoskeleton and bilayer surface, nor
does it account for inextensibility of spectrin beyond its natural
contour length; the individual spectrin links are treated as
simple Gaussian chains. Individual links in the network are
allowed to break and reform following kinetic equations that do
not obey detailed balance; this provides a model for energy (ATP)
consumption at the membrane surface. $A$ is the average mesh size
($\sim 100 nm$).} \label{fig:snap}
\end{figure}

In this paper, we extend the DF entropic spring model of the
cytoskeleton meshwork to include dynamic evolution of the
entire system.  We allow the anchor points between spectrin and
membrane to laterally diffuse and we allow for dynamic dissociation
and association of spectrin links as a molecular level manifestation
of the non-thermal energetic picture proposed by GS (fig. \ref{fig:snap}).

\begin{figure}
\begin{center}
\includegraphics[width=0.5\textwidth]{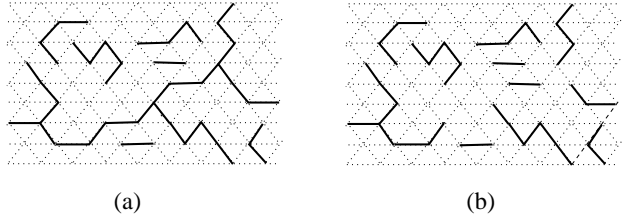}
\end{center}
\caption{A schematic illustration of a randomly broken spectrin
meshwork. The solid lines are connected links while the dotted
lines represent disconnected ones.  The percentage of intact links
(equivalently, the probability for any single link to be intact)
will be denoted ``$p$''.  All lines together (solid and dotted)
correspond to a complete hexagonal meshwork. (a): Above the
percolation threshold, $p>p_c$, the connected links have global
connectivity, i.e., they form a cluster extending across the
entire cell. The corresponding effective surface tension will
always be finite. (b): Below the percolation threshold, $p<p_c$,
there is no global connectivity in the meshwork, i.e. connected
links form finite islands of connectivity that do not span the
cell. The corresponding effective surface tension becomes zero if
the links are breaking and reforming slowly. (c): Even at
instantaneous connectivities below the percolation threshold,
there can be a finite surface tension if spectrin
breaking/formation kinetics are sufficiently fast. In this limit,
each link can be thought of as a spring with a reduced restoring
force since out of plane membrane undulations evolve more slowly
than the lifetime of the link.} \label{fig:perco}
\end{figure}

One important consequence
of the GS picture is that sufficiently high ATP concentrations lead to an
appreciable fraction of dissociated spectrin links at the membrane surface.
Depending upon the timescales for spectrin (re)association, the effective
long-wavelength elastic
properties of the membrane interpolate between two limiting cases.  If
spectrin (re)association kinetics are much slower than all other timescales
in the problem, the effective tension imposed by the network (as inferred by
out-of-plane bilayer undulations) is well predicted by
a simple percolation-theory argument (see fig. \ref{fig:perco}).  In the opposite limit of fast spectrin
kinetics, the effective tension is well predicted by assuming each link in
the network has a reduced spring constant proportional to the probability
of the link being intact at steady state.  At intermediate rates, simulations
are seen to interpolate between the two extreme cases.


This paper is organized as follows. In Sec.~\ref{sec:model}, we
present our mathematical model for the RBC membrane. In
Sec.~\ref{sec:simulation} details of our simulation methods are
discussed. In Sec.~\ref{sec:complete} results for a fully intact
spectrin meshwork are presented, while in Sec.~\ref{sec:broken} we
generalize to the more interesting case of a randomly broken
network (both static and dynamically broken). We discuss our
results in relation to experiment in Sec.~\ref{sec:experiments}
and conclude in Sec.~\ref{sec:conclusion}.

\section{Model}
\label{sec:model}

We treat the RBC membrane as a Helfrich fluid sheet \cite{Helfrich}
coupled via mobile anchor points to a network of springs.  Our Hamiltonian
is
\begin{eqnarray}
H&=&\int_{\mathcal{A}}
d\mathbf{r}\frac{\kappa}{2}[\nabla^2h(\mathbf{r})]^2\nonumber\\&+&
\sum_{\langle i,j\rangle}\xi_{ij}(t)\left\{\frac{\mu}{2}
(\mathbf{r}_i-\mathbf{r}_j)^2
+\frac{\mu}{2}[h(\mathbf{r}_i)-h(\mathbf{r}_j)]^2+E\right\}.\nonumber\\\label{eq:H}
\end{eqnarray}
The first term (integral portion) is the standard bilayer bending
energy for a Monge gauge sheet assuming small deformations~\cite{Helfrich} and
 bending modulus $\kappa$. $\mathcal{A}$
is the projected area of the membrane, $\mathbf{r}=(x,y)$ is the
position vector in the $xy$ plane and and $h(\mathbf{r})$ is the
local displacement of the membrane away from the flat reference
configuration specified by $h(\mathbf{r})=0$ (see
Fig.~\ref{fig:snap}).  We assume that the lipid bilayer itself has
a negligible (bare) surface tension.  In more general situations,
eq. \ref{eq:H} is easily modified to account for non-vanishing
tension inherent to the lipid bilayer portion of the membrane
\cite{Safran,Helfrich}. The second term (sum portion) accounts for
the energy of the 2D cytoskeletal meshwork, modelled as a network
of entropic springs (or in other words, ideal chains of polymers)
with effective spring constant $\mu$ ($\mu = 3k_B T/\ell_k\ell_c$
with $\ell_k$ and $\ell_c$ the Kuhn and contour length of the
polymer, respectively). In a fully intact cytoskeletal meshwork,
all spring end points (nodes) are restricted to lie on the surface
of the bilayer, so their coordinates are specified by
$\mathbf{r_i}$ and $h(\mathbf{r_i})$, with indices $i$ (or $j$)
labelling different nodes. The sum is over all \emph{distinct}
nearest neighbor node pairs (equivalently, over all polymer
springs), denoted as $\langle i,j\rangle$. The factor
$\xi_{ij}(t)$ is included to account for the possibly incomplete
connectivity of the network. It is equal to 1 when the $ij$ link
is connected and 0 otherwise. The constant $E$ reflects all other
free energy change associated with connecting a detached filament
end to a node besides the elastic energy of the spring (e.g., the
binding energy to the node); we assume this energy is negative and
significantly larger than thermal energy scales to insure
stability of the network in the absence of non-thermal energy
sources.

Although our starting point is very similar to the DF model, we emphasize
a few key differences.  In DF, the lateral positions of nodes are held fixed in the
geometry of minimum energy for a flat bilayer surface;  {\em i.e.} the $r_i$ variables
are treated as set constants in DF, not as variables capable of influencing
the energetics and/or dynamics of the system.  This approximation renders  the
Hamiltonian analytically tractable, however it is not immediately clear that such
a choice fully captures all relevant physics in this system.
For example, with fixed node positions equally spaced on a regular lattice, the
``spring'' contribution to eq. \ref{eq:H} amounts to a finite differenced version of the
usual surface tension contribution to the Helfrich Hamiltonian.  At long wavelengths,
eq. \ref{eq:H} with fixed nodes is guaranteed to behave as a Helfrich sheet under tension.
If the nodes are mobile, as physically expected, the spring network represents a simple manifestation
a tethered membrane.  Such membranes are known to exhibit more complicated fluctuations
than expected for Helfrich fluid bilayers \cite{nelson}.  We also emphasize that
much of the following work is concerned with membrane dynamics, which was not
considered in DF.  One of the most interesting aspects of our model is the dynamic
breaking and reformation of cytoskeletal filaments, which could not be studied with
the equilibrium approach adopted by DF.

Dynamics in our system are overdamped, owing to the low Reynolds
number environment present at cellular length scales \cite{purcell}.
For bilayer height fluctuations, we have the following Langevin
type equation of motion, which accounts for hydrodynamic flow
in the surrounding cytoplasm  ~\cite{Granek,Lin,brown_accounts}
\begin{equation}
\frac{\partial h(\mathbf{r},t)}{\partial t}=\int_{-\infty}^{\infty}
d\mathbf{r}^{\prime}\Lambda(\mathbf{r}-\mathbf{r}^\prime)
[F(\mathbf{r}^\prime,t)+\zeta(\mathbf{r}^\prime,t)].\label{eq:langevin}
\end{equation}
Here, $\Lambda(\mathbf{r})$ is the diagonal part of the Oseen
tensor~\cite{Edwards}, given by
\begin{equation}
\Lambda(\mathbf{r})=\frac{1}{8\pi\eta|\mathbf{r}|},
\end{equation}
where $\eta$ is the viscosity of the surrounding fluid.  The
above integral is taken over the entire $x,y$ plane; it is assumed that
the area of interest, $\mathcal{A}$, is embedded within a periodically
repeating environment of identical subsystems.
$F(\mathbf{r},t)$ is the force per unit area on the membrane,
\begin{eqnarray}
F(\mathbf{r},t)&=&-\frac{\delta H}{\delta h(\mathbf{r},t)}
=-\kappa\nabla^4h(\mathbf{r},t)\nonumber\\&-&\sum_{\langle
j\rangle}\mu
\xi_{ij}(t)\delta^2(\mathbf{r-r}_i)[h(\mathbf{r},t)-h(\mathbf{r}_j,t)],
\label{eq:Fr}
\end{eqnarray}
where $\delta(\mathbf{r})$ is the Dirac delta function, and the
sum is over all nearest neighbors of node $i$, denoted as $\langle
j\rangle$. $\zeta(\mathbf{r},t)$ is a spatially local Gaussian
white noise satisfying the fluctuation-dissipation relation
\begin{eqnarray}
\langle\zeta(\mathbf{r},t)\rangle &=&0,\\
\langle\zeta(\mathbf{r},t)\zeta(\mathbf{r}^\prime,t^\prime)\rangle&=&
2k_BT\Lambda^{-1}(\mathbf{r-r^\prime})\delta(t-t^\prime),
\end{eqnarray}
where $k_B$ is Boltzmann's constant and $T$ temperature
of the system.

We have another set of Langevin equations describing lateral diffusion
of the nodes within the bilayer
\begin{equation}
\frac{\partial \mathbf{r}_i(t)}{\partial t}=\frac{D}{k_BT}[
\mathbf{F}_i(t)+\boldsymbol{\zeta}_i(t)],\label{eq:langevin2}
\end{equation}
with
\begin{eqnarray}
\mathbf{F}_i(t)&=&-\frac{\partial H}{\partial
\mathbf{r}_i}\nonumber\\&=&-\sum_{\langle j\rangle}\mu
\xi_{ij}(t)\left \{(\mathbf{r}_i-\mathbf{r}_j)
+[h(\mathbf{r}_i)-h(\mathbf{r}_j)]\frac{\partial
h(\mathbf{r}_i)}{\partial \mathbf{r}_i}\right\},\nonumber\\
\label{eq:Fir}
\end{eqnarray}
and
\begin{eqnarray}
\langle\boldsymbol{\zeta}_i(t)\rangle &=&0,\\
\langle\boldsymbol{\zeta}_i(t)\cdot\boldsymbol{\zeta}_j(t^\prime)\rangle&=&
\frac{4(k_BT)^2}{D}\delta_{ij}\delta(t-t^\prime),
\end{eqnarray}
where $D$ is the lateral diffusion constant of the node across the
membrane surface. Eqs. \ref{eq:langevin} and \ref{eq:langevin2}
completely specify the dynamics of the lipid bilayer and the attached 2D meshwork.
Notice that the two sets of Langevin equations are coupled (and must be solved
simultaneously) via the shape of the membrane surface.
We note that eq. \ref{eq:Fir} neglects the purely geometric effect of non-flat membrane
geometry on the $(x,y)$ motion of node points \cite{seifertD,ali}.  This approximation
significantly simplifies our modeling and it has recently been demonstrated that such
geometric effects are very small for the physical parameters studied herein \cite{ali}.

It is convenient to recast eq.~(\ref{eq:langevin}) in a Fourier
basis ~\cite{Lin}.
\begin{equation}
\frac{\partial h_{\mathbf{k}}(t)}{\partial
t}=\Lambda_{\mathbf{k}}[F_{\mathbf{k}}(t)+\zeta_{\mathbf{k}}(t)]\label{eq:langevink}
\end{equation}
with $\mathbf{k}=(2\pi m/L_x,2\pi n/L_y)$ for integer $m$ and $n$.
Here for the latter convenience of the simulation, we assume in
general a rectangular sample with size $L_x\times L_y$ in real
space, and therefore the two lattice constants in $k$ space are
different. The quantities $h_{\mathbf{k}}$, $F_{\mathbf{k}}$ and
$\zeta_\mathbf{k}$ derive from functions periodic in $x$ and $y$,
due to the assumed periodicity of the system.  The Fourier
transform pair for an arbitrary function, $f$, with such
periodicity is
\begin{eqnarray}
f(\mathbf{r})=\frac{1}{L_xL_y}\sum_\mathbf{k}f_\mathbf{k}e^{i\mathbf{k}\cdot
\mathbf{r}},\label{eq:Fourierk}\\
f_\mathbf{k}=\int_{\mathcal{A}}
d\mathbf{r}f(\mathbf{r})e^{-i\mathbf{k}\cdot\mathbf{r}}.\label{eq:Fourierr}
\end{eqnarray}
The Fourier transformed Oseen interaction,
\begin{equation}
\Lambda_{\mathbf{k}}=\int_{-\infty}^{\infty}d\mathbf{r}e^{-i\mathbf{k}\cdot\mathbf{r}}
\Lambda(\mathbf{r})=\frac{1}{4\eta k},
\end{equation}
in contrast, reflects transformation over the full 2D plane.  By
construction, the dynamics specified by eq. \ref{eq:langevink}
reflects an infinite network of periodic membrane images
interacting via the long range $1/r$ Oseen hydrodynamic kernel.
The random forces obey
\begin{eqnarray}
\langle\zeta_{\mathbf{k}}(t)\rangle &=&0,\\
\langle\zeta_{\mathbf{k}}(t)\zeta_{\mathbf{k}^\prime}(t^\prime)\rangle&=&
2k_BTL_xL_y\Lambda_{\mathbf{k}}^{-1}\delta_{\mathbf{k},-\mathbf{k}^{\prime}}
\delta(t-t^\prime).
\end{eqnarray}

For the dynamics of the breaking and reforming of spectrin springs,
we consider a simple two state kinetic model.  We define the rate to reconnect
a link as $k_c$, and the rate to disconnect a link as $k_d$, irrespective
of the instantaneous position of the endpoints of the spring.
Accordingly, the value of each $\xi_{ij}$ jumps back and
forth between 1 and 0. Defining $p$ as the steady-state probability of a
link to be connected at any moment, we have
\begin{equation}
p=\frac{k_c}{k_c+k_d}.\label{eq:pk}
\end{equation}
It should be emphasized that our simple model for the breaking
and reformation of spectrin filaments does not obey detailed balance
since we do not account for the variations in energetics caused by
positions of the network nodes within the kinetic scheme.  This approach
necessarily corresponds to a non-equilibrium situation, with the
dynamics of the spectrin network driving membrane fluctuations in
a non-thermal manner.  Qualitatively, this corresponds to the picture
proposed by GS, however the detailed kinetics involved in the spectrin
(re)association process are unknown and likely differ substantially from
the picture adopted herein.  Our simple two-state picture
represents one possible manifestation of non-equilibrium driving.

The coupled equations implied by eqs. \ref{eq:langevink} and
\ref{eq:langevin2} are not amenable to analytical solution and
will be solved via simulation as detailed below.  Before proceeding,
we note that simulations are run on a discrete square
lattice. In other words, Eq.~(\ref{eq:Fourierr}) is replaced by
\begin{equation}
f_\mathbf{k}=\sum_\mathbf{r}
a^2f(\mathbf{r})e^{-i\mathbf{k}\cdot\mathbf{r}},
\end{equation}
where $a$ is the lattice constant and $\mathbf{r}$ now only take
discrete values on the lattice. Correspondingly, in $\mathbf{k}$
space, the reciprocal lattice (with lattice constant $2\pi/L_x$
and $2\pi/L_y$) contains $(L_x/a)\times(L_y/a)$ points and the
summation in eq. \ref{eq:Fourierk} is finite.

\begin{figure}
\begin{center}
\includegraphics[width=0.5\textwidth]{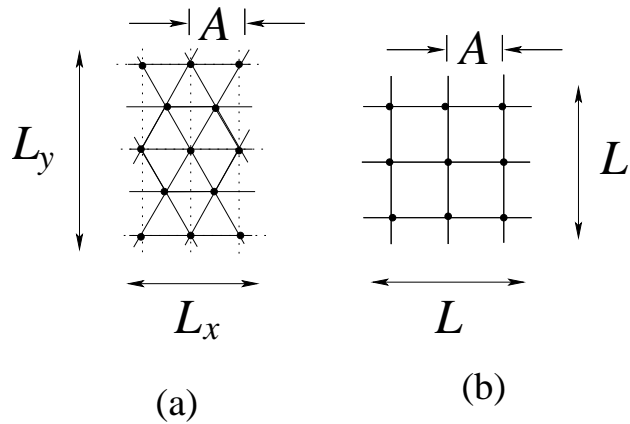}
\end{center}
\caption{A schematic illustration of the connectivity of the nodes
of the meshwork. (a): the hexagonal meshwork. The dotted lines are
a guide of eyes for the rectangular sample used. (b): the square
meshwork.} \label{fig:corral}
\end{figure}

Two types of meshworks are considered in our simulations. First,
in accordance with the true geometry of RBC membranes~\cite{Boal},
we consider a hexagonal meshwork (Fig.~\ref{fig:corral}a). In
such a meshwork, when all links are connected, the average
positions of all nodes form a hexagonal lattice (this lattice of
nodes should not be confused with the lattice used to discretize
the surface as discussed above). We consider a finite sample and
assume periodic boundary conditions in a rectangular geometry
of size $L_x\times L_y$ approximately commensurate with the embedded
hexagonal meshwork (see Fig.~\ref{fig:corral}a).
The average distance between the nearest neighbor nodes, $A$, is
determined from the average surface density of nodes in a RBC
membrane. For theoretical interest, we also consider a square
meshwork with square lattice symmetry for the average
positions of nodes (Fig.~\ref{fig:corral}b). Here we simply take a
periodic square box, i.e., $L_x=L_y=L$.

To connect with experiment and prior theoretical work, our simulations
are used primarily to calculate the mean square height
fluctuation of the membrane surface in $\mathbf{k}$ space, $\langle
|h_\mathbf{k}|^2\rangle$ (angular brackets represent non-equilibrium averages
as well as equilibrium averages in this work). At long wavelengths ($\lambda \gg A$)
we observe that the relation
\begin{equation}
\langle |h_\mathbf{k}|^2\rangle=\frac{k_BTL_xL_y}{\kappa_{eff}
k^4+\sigma_{eff} k^2}.\label{eq:hk2}
\end{equation}
holds fairly well.  This expression corresponds to that expected \cite{Helfrich,Safran}
for a thermal fluid bilayer sheet with effective bending rigidity $\kappa_{eff}$ and
effective surface tension $\sigma_{eff}$, but we stress that its use in interpreting the simulations
is empirical.  The composite membrane surfaces studied in this work can not be regarded
as fluid-like due to the assumed connectivity of the cytoskeleton matrix.  Much of our analysis is presented in terms
of $\sigma_{eff}$, which is obtained by fitting the simulated data to eq.
\ref{eq:hk2} (while assuming $\kappa_{eff}=\kappa$ unless noted otherwise).
For future reference, we define the ``free fluctuation spectrum"
of the sheet, $\langle|h_\mathbf{k}|^2\rangle_f$, as the result anticipated from
eq. \ref{eq:H} in the absence of any cytoskeletal contributions (i.e. all $\xi_{ij}=0$)
\begin{equation}
\label{eq:free}
\langle |h_\mathbf{k}|^2\rangle_f=\frac{k_BTL_xL_y}{\kappa k^4}.
\end{equation}

\section{simulation methods}\label{sec:simulation}

Two simulation methods were used to study the composite membrane
system: Fourier Monte Carlo (FMC)~\cite{Nagle,Nagle2} and Fourier
space Brownian dynamics (FSBD)~\cite{LinBJ,LinPRL,Lin}. In the
limit of infinitely slow breaking and reformation of spectrin
filaments (quenched disorder), the $\xi_{ij}$ variables are static
over the course of any finite simulation and the system relaxes to
a thermal equilibrium dependent upon the connectivity of the
network.  In this limit, both FMC and FSBD simulations may be used
to calculate thermal averages and both should agree with one
another. When spectrin links are breaking and reforming in time
following the non-equilibrium scheme presented above, we must run
dynamic FSBD calculations.  Our primary interest in this work is
the non-equilibrium case, however FMC calculations were performed
both to verify the accuracy of our FSBD simulations (in the static
network limit) and to examine the scaling of some our equilibrium
results with system size (the FMC method is computationally more
efficient than FSBD).

\subsection{Fourier Monte Carlo (FMC)}

The FMC scheme \cite{Nagle,Nagle2} is a standard Metropolis
algorithm \cite{Frenkel}, which uses Fourier modes of the bilayer,
$h_{\mathbf{k}}$, as the bilayer degrees of freedom. There are two
kinds of degrees of freedom in our system: membrane shape modes
and lateral position of the nodes of the spectrin network. For the
former,  we attempt MC moves on the Fourier modes of the system to
enhance computational efficiency relative to naive real space
schemes~\cite{Nagle,Nagle2}. While for the latter, we attempt
moves that displace the $x,y$ position of the nodes to adjacent
sites of the real space lattice inherent to the simulations.  Note
that the shape of the membrane surface is continuously variable in
our description, while the lateral position of nodes is discrete.
The primary advantage to evolving the shape of the bilayer surface in
Fourier space is that we may tune the maximal size of attempted MC
moves for each mode separately.  In the case of a simple fluid bilayer
(without attached cytoskeleton) under finite surface tension it is clear
that a good choice for maximal move sizes is
(see Eq.~(1) in Ref.~\cite{Bouzida})
\begin{equation}
\frac{(\kappa k^4+\sigma k^2)\delta_\mathbf{k}^2}{2L_xL_y}=const.,
\end{equation}
where $\delta_\mathbf{k}$ is the maximum attempted jump size of mode
$\mathbf{k}$.  In our case, a similar choice works well only for wavelengths
sufficiently long that eq. \ref{eq:hk2} is obeyed, however it is a simple
matter to tune each $\delta_\mathbf{k}$ individually to optimize simulation
efficiency.  In practice, maximal jump sizes were tuned to insure that the acceptance
ratio of all trials was approximately $1/2$.

\subsection{Fourier space Brownian dynamics (FSBD)}

The FSBD method has been fully detailed in our previous work
\cite{LinBJ,LinPRL,Lin}. Here we present only a minimal discussion
to introduce the method.  In this section we consider the
simple case when $p_{ij}$ in Eq.~(\ref{eq:H}) is independent of
time and postpone the discussion of $p_{ij}(t)$ for the next
subsection.

Integrating Eqs.~(\ref{eq:langevink}) and~(\ref{eq:langevin2})
from $t$ to $t+\Delta t$ for small $\Delta t$, we have
\begin{eqnarray}
h_{\mathbf{k}}(t+\Delta
t)=h_{\mathbf{k}}(t)+\Lambda_{\mathbf{k}}[F_{\mathbf{k}}(t)\Delta
t+R_{\mathbf{k}}(\Delta t)],\nonumber\\
\mathbf{r}_i(t+\Delta
t)=\mathbf{r}_i(t)+\frac{D}{k_BT}[\mathbf{F}_i(t)\Delta
t+\mathbf{R}_i(\Delta t)].\label{eq:fsbd}
\end{eqnarray}
where
\begin{equation}
R_{\mathbf{k}}(\Delta t)=\int^{t+\Delta t}_t
dt^\prime\zeta_{\mathbf{k}}(t^\prime), \     \ \mathbf{R}_i(\Delta
t)=\int^{t+\Delta t}_t dt^\prime\boldsymbol{\zeta}_i(t^\prime).
\end{equation}
The statistical properties of $R_{\mathbf{k}}(\Delta t)$ and
$\mathbf{R}_i(\Delta t)$ follow directly those of
$\zeta_{\mathbf{k}}(t)$ and $\boldsymbol{\zeta}_i(t)$,
\begin{eqnarray}
\langle R_{\mathbf{k}}(\Delta t) \rangle&=&0,\\ \langle
R_{\mathbf{k}}(\Delta t)R_{\mathbf{k}^\prime}(\Delta t)
\rangle&=&2k_BTL^2\Delta
t\Lambda_{\mathbf{k}}\delta_{\mathbf{k},-\mathbf{k}^\prime},\label{eq:Rk}\\
\langle \mathbf{R}_i(\Delta t) \rangle&=&0,\\ \langle
\mathbf{R}_i(\Delta t)\cdot\mathbf{R}_j(\Delta t)
\rangle&=&4(k_BT)^2\Delta t/D\delta_{i,j}.
\end{eqnarray}
In the simulation, $R_{\mathbf{k}}(\Delta t)$ and
$\mathbf{R}_i(\Delta t)$ are drawn from Gaussian distributions
with means and variances specified above.
Since $h_{\mathbf{k}}=h_{\mathbf{-k}}^\ast$ must be satisfied to
insure the height field of the membrane is real valued, only about
half of the $R_k$ need to be generated in each time step; the
remaining follow via complex conjugation.  The precise formulation
of this statement is somewhat complicated by the finite number of
modes in our discrete Fourier transform.  Readers are referred to
ref. \cite{LinJCTC} for a detailed discussion of how the full set
of random forces are to be generated while preserving the real
valued nature of $h(\mathbf{r})$.

At each time step of the FSBD simulation, the following
calculations are performed:

(1) Take $h(\mathbf{r},t)$ and $\mathbf{r}_i(t)$ from the last
time step.

(2) Evaluate $F(\mathbf{r},t)$ and $\mathbf{F}_i(t)$ using
Eqs.~(\ref{eq:Fr}) and~(\ref{eq:Fir}).

(3) Compute $F_{\mathbf{k}}(t)$ by Fourier transforming
$F(\mathbf{r},t)$.

(4) Draw $R_\mathbf{k}(\Delta t)$ and $\mathbf{R}_i(\Delta t)$
from the Gaussian distributions specified above.

(5) Compute $h_{\mathbf{k}}(t+\Delta t)$ and
$\mathbf{r}_i(t+\Delta t)$ using Eq.~(\ref{eq:fsbd}).

(6) Get $h(\mathbf{r},t+\Delta t)$ through inverse Fourier
transformation for use in the next iteration.

There is one complication in step 2 of the above procedure.  While
we evolve each $\mathbf{r}_i (t)$ as a continuous variable, the
height field of the membrane is only specified at points on the
real space lattice (more precisely, it is only readily obtainable
via Fast Fourier Transformation at these points).  To compute the
forces for use in eqs. \ref{eq:Fr} and \ref{eq:Fir}, both
$\mathbf{r}_i$ and $h(\mathbf{r}_i)$ are approximated by assuming
the position of node $i$ directly coincides with the nearest real
space lattice site.  While this approximation could potentially
cause problems due to discontinuous jumps in the forces, the
surfaces under study are only weakly ruffled.  It was verified (in
the thermal case) that FMC simulations with node positions
strictly confined to lattice sites and FSBD simulations as
outlined here were in good agreement.   In practice, we choose
$\Delta t$ small enough that further reduction has no consequence
for the reported results, typically on the order of $0.01$ $\mu$s
depending upon choice of lattice size $a$.

\subsection{Kinetic Monte Carlo (KMC) for spectrin (re)association kinetics}

For the case of a dynamic network, every $\xi_{ij}(t)$ jumps back and forth
between 1 and 0, according to the two rates $k_c$ and $k_d$ (see
Sec.~\ref{sec:model}). We use the stochastic simulation algorithm of
Gillespie (kinetic Monte Carlo)~\cite{Gillespie} to pick random times
for breaking and reformation events to occur.
Let us consider the event of reconnecting a link. If the link is
disconnected at time $t=0$, then the waiting time distribution
for link reconnection is $W(t)=k_c e^{-k_c t}$ \cite{Vankampen}.
A random number consistent with this distribution is obtained
by applying the following transformation to a uniform random
deviate $x$ \cite{NumRec}
\begin{equation}
t=-\frac{1}{k_c}\ln x.
\end{equation}
A similar transformation, replacing $k_c$ with $k_d$ is used
to determine the time at which an intact filament breaks apart.
The set of times so obtained provides a complete trajectory for
the behavior of each $\xi_{ij}$ for use in eqs. \ref{eq:Fr} and
\ref{eq:Fir}.  The continuously distributed times are rounded off
to the nearest time point in the discrete FSBD procedure.
We note that our model assumes each link must
always connect the same two nodes (or be broken) - there
is no provision for a spectrin filament to dissociate from one
node and reconnect elsewhere.

\section{Effective surface tension in the presence of an intact cytoskeletal meshwork}
\label{sec:complete}

In this section we assume a fully connected cytoskeleton meshwork
($\xi_{ij} = 1$ for all links at all times). In this limit, there are no non-thermal
effects and either KMC  or FSBD
simulations may be performed to calculate the equilibrium spectrum,
$\langle|h_\mathbf{k}|^2\rangle$.  The qualitative features of
this fluctuation spectrum have been predicted by Fournier et. al.
\cite{FournierPRL}.  They argued that the behavior of the
composite membrane surface should behave simply in two limits.  In
the short wavelength limit, the effect of the cytoskeleton might
be expected to play a minor role; neglecting the cytoskeletal
terms in eq. \ref{eq:H} leads to the prediction
$\langle|h_\mathbf{k}|^2\rangle = k_B T L_xL_y/\kappa k^4$. In the
long wavelength limit, the cytoskeleton should play an important
role, but may be regarded as a continuous medium imparting an
effective surface tension to the bilayer (and possibly modifying
the bare bilayer bending rigidity) as in eq. \ref{eq:hk2}.
``Short'' and ``long'' wavelengths referenced above are understood
to be interpreted relative to the size of the individual spectrin
links ($A$, see fig \ref{fig:corral}).  The original work of
Fournier et. al. assumed a sharp transition between these two
regimes and fit experimental data with a transition at a wavelength of approximately
$2\pi A$. The simulations below are in qualitative agreement with
this picture, but place the transition wavelength at $A$ and
predict a finite width to the crossover region.

Both FMC and FSBD simulations were
performed with identical results (as expected).  Simulations were
seeded from an initially flat membrane with the cytoskeletal anchor points
arranged in a perfect lattice (as indicated in fig. \ref{fig:corral}).  The
initial configuration was allowed to fully equilibrate before collecting any
data for analysis.  The real-space lattice constant used in the simulations
was $a=A/4$ with box dimensions of $L_x=32a$ and $L_y=42a$ for the
hexagonal network and $L_x=L_y=L=32a$ for the square network (except where
indicated otherwise, these values of $a$ and $L_x$,$L_y$ and $L$ were used
in all reported simulations).  This
corresponds to 96 independent nodes (192 triangular corrals within the periodic box) in the case of six-fold
connected anchors and 64 independent nodes (64 square corrals within the periodic box) in the case
of the four-fold connected anchors.  In preliminary runs, it was
verified that neither increasing the sample size ($L_x$,$L_y$) nor decreasing
$a$ by a factors of $2$ significantly altered results; the values outlined
above were thus chosen to insure converged results with minimal computational
expense.
For convenience, all physical parameters used in the simulations are listed in
Table~\ref{tab:para}.



\begin{figure}
\begin{center}
\includegraphics[width=0.45\textwidth]{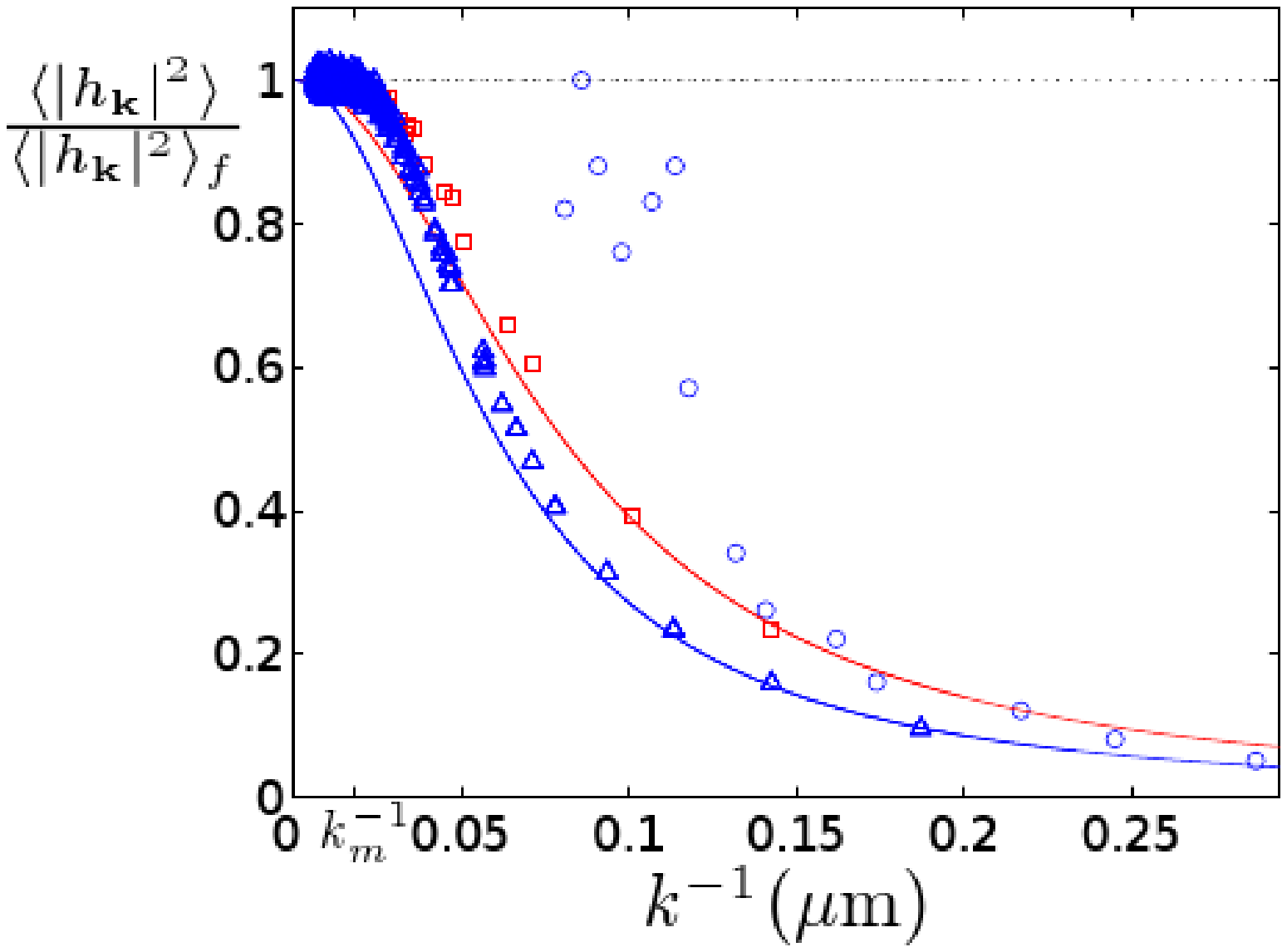}
\includegraphics[width=0.45\textwidth]{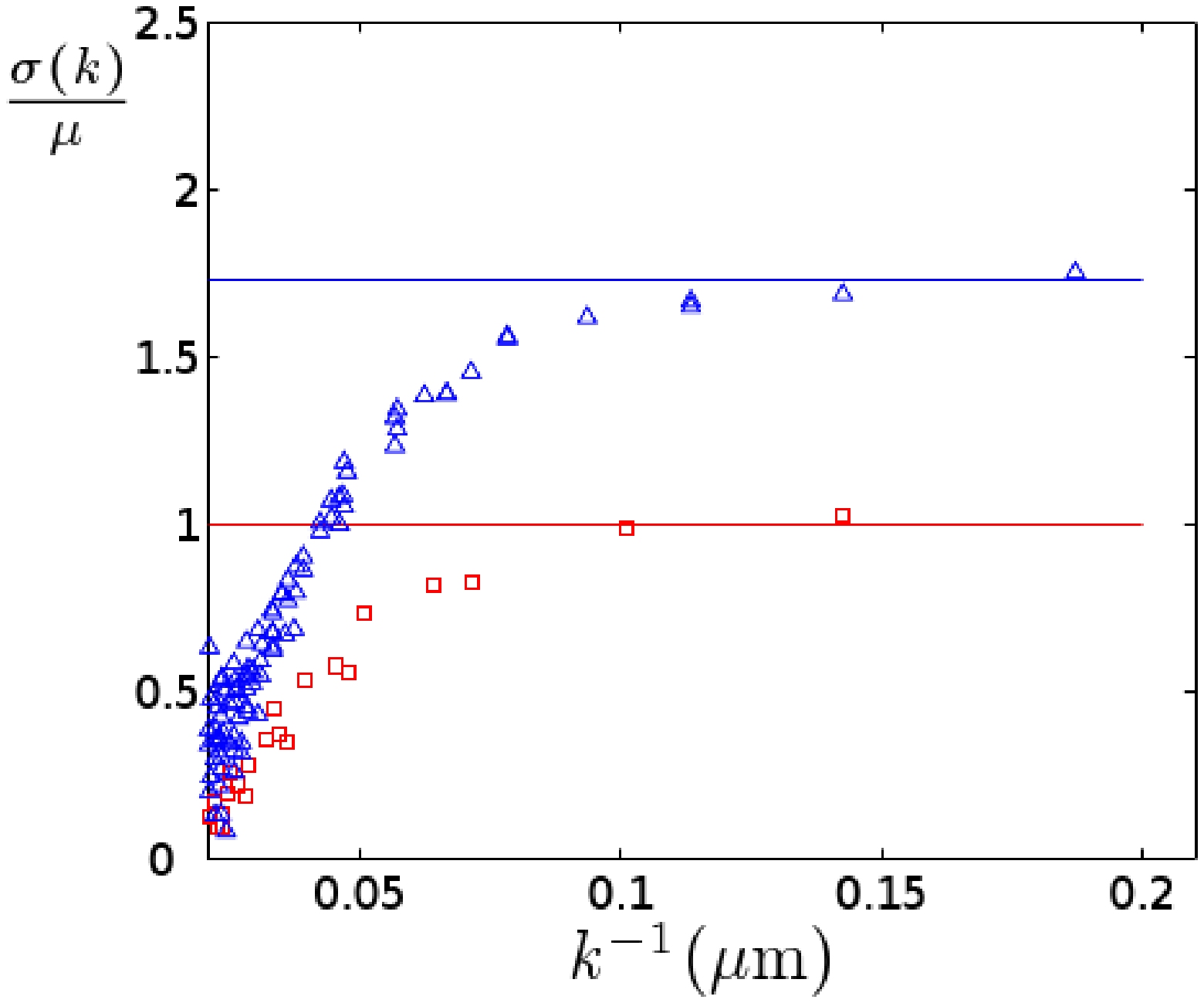}
\end{center}
\caption{UPPER: Spectrum of fluctuations,
$\langle|h_{\mathbf{k}}|^2\rangle$, (normalized by
$\langle|h_{\mathbf{k}}|^2\rangle_f$) versus inverse wavenumber
for the our RBC membrane model assuming a perfectly intact
cytoskeletal network. The squares and triangles indicate
simulation results for square and hexagonal meshworks
respectively. The error bars are approximately the same size as
the symbols. The solid lines are theoretical results using
Eq.~(\ref{eq:sigma}) for  $\sigma_{eff}$ and $\kappa_{eff}=\kappa$
in Eq. \ref{eq:hk2}.  The wavevector $k_m=2\pi / A$ is indicated
on the horizontal axis to show where crossover from free membrane
to long-wavelength behavior might be expected to occur.  The
circles are experimental results taken from Ref.~\cite{Zilker}.
LOWER: The approach of $\sigma$ to its limiting long-wavelength
value is displayed.  In this case, each point in the spectrum on
the left was individually fit to eq. \ref{eq:hk2} by adjusting
$\sigma_{eff}$ (assuming $\kappa_{eff}=\kappa$).  The resulting
$\sigma_{eff}$ as a function of $k^{-1}$ are shown.  At the
longest wavelengths simulated, the inferred tensions converge to
the predicted values.} \label{fig:mcmesh}
\end{figure}

Our simulation results are shown in Fig.~\ref{fig:mcmesh}.
In the long wavelength limit, the results are well fit by eq.
\ref{eq:hk2}, assuming $\kappa_{eff} = \kappa$ and using
\begin{equation}
\sigma_s=\mu, \ \ \sigma_h=\sqrt{3}\mu.\label{eq:sigma}
\end{equation}
for the effective tension in the square and hexagonal symmetry
simulations, respectively.  These tension values are not fit constants,
but rather may be inferred from the cytoskeletal contribution to
eq. \ref{eq:H}.  Surface tension is an energy per unit area, so we may
calculate its value as the ratio of entropic spring (cytoskeleton) energy
per corral to the area per corral.  In the square geometry, there are effectively
two springs per corral (each spring is shared by two adjacent corrals)
and using an idealized zero temperature geometry, a single corral has area $A^2$ and
total spring energy of $2\times\frac{\mu}{2}A^2$.  The reported value for $\sigma_s$
follows immediately.  A similar calculation leads to the somewhat larger value
of $\sigma_h$ in the hexagonal geometry, reflecting the higher density of springs
in this case.  The numerical value of $\sigma_h$ so calculated is $1.3\times 10^3$
k$_\mathrm{B}$T $\mu$m$^{-2}=5.3 \times 10^{-6}$ J m$^{-2}$, which
is close to a theoretical fit~\cite{FournierPRL} of the
experimental result~\cite{Zilker} (see Fig.~\ref{fig:mcmesh}).

The fluctuation spectra in fig. \ref{fig:mcmesh} indicate that free membrane
predictions hold reasonably well out to wavelengths of approximately $A$ ($k^{-1} = A/(2\pi) \sim 0.02\mu \mathrm{m}$), with
good convergence to long-wavelength behavior established by $5A$ ($k^{-1} \sim 0.1 \mu \mathrm{m}$).  The
intermediate regime between wavelengths of $A$ to $5A$ encompasses the crossover between the
two limiting cases.  While this fact is unfortunate in light of the experimental data for the RBC \cite{Zilker},
which displays a crossover
at longer wavelengths, the simulation predictions are unambiguous.  The experimental data remains a mystery,
but we do note that it may be accounted for by adoption of an ad hoc harmonic confining potential \cite{GovPRL}.

\begin{figure}
\begin{center}
\includegraphics[width=0.5\textwidth]{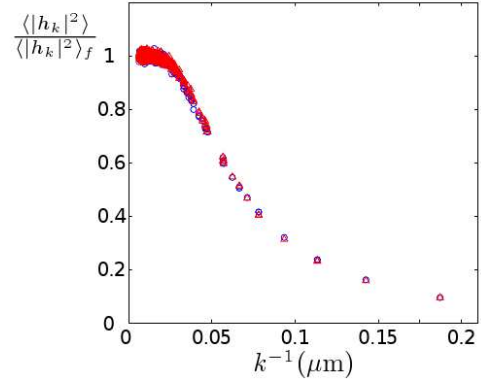}
\end{center}
\caption{Simulation results of $\langle|h_\mathbf{k}|^2\rangle$
for a hexagonal lattice. The circles are the data for the case of
immobile nodes placed on a perfect lattice. While the triangles
are the data for the case of mobile nodes (identical data to fig.
\ref{fig:mcmesh}).  Mobility of the cytoskeleton attachment points
does not significantly alter the simulation results.}
\label{fig:pineornot}
\end{figure}

\begin{figure}
\begin{center}
\includegraphics[width=0.5\textwidth]{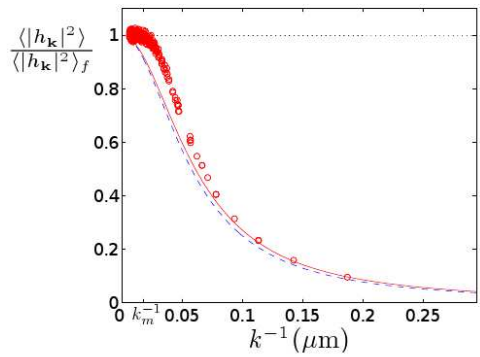}
\end{center}
\caption{Comparison of different theories for a hexagonal lattice.
The circles are the simulation results. The dashed line is the
prediction introduced in fig. \ref{fig:mcmesh}. The solid line is
the DF model prediction.  At intermediate wavelengths DF does a
slightly superior job in fitting the simulations.  Both
approximations fail at small wavelengths and converge to the same
behavior at long wavelengths.} \label{fig:kccorrect}
\end{figure}

One of the motivations for this work was to test the performance of the analytical theory
developed in DF.  In particular, we anticipated that the mobility of cytoskeletal anchors would lead to some
quantitative deviations from the DF theory at short wavelengths.  In fact, the mobility of anchor points leads
to insignificant changes in the fluctuation spectrum for physical parameters relevant to the RBC (see fig. \ref{fig:pineornot}).  We
also note that the detailed analysis of DF does slightly better in reproducing the simulated spectrum than does
the adoption of surface tensions implied by eq. \ref{eq:sigma} (see fig. \ref{fig:kccorrect}).  At intermediate
wavelengths, the small deviations of $\kappa_{eff}$ away from $\kappa$ predicted by DF do improve the fit as compared
to our naive arguments, however the effect is very slight in comparison to the leading order
effect of introducing a finite surface tension.
As a final point, we note that the anisotropic nature of the square network over all wavelengths
leads to some variance in fluctuations for a given magnitude of $k$ depending on the direction taken.
In principle, the hexagonal network should not suffer from this effect \cite{Boal}, however the underlying square lattice taken for our simulations does introduce some anisotropy to the spectra; these effects are most severe at short wavelengths.
The spread of data points plotted in figs. \ref{fig:mcmesh} - \ref{fig:kccorrect}  reflects this directional dependence in
the spectra and should not be taken as evidence of statistical noise.  As previously mentioned, statistical errors are
less than the size of the symbols used in plotting.



\section{Effective surface tension in the presence of a cytoskeletal meshwork with dynamically evolving connectivity}
\label{sec:broken}

We now generalize the results of the previous section
to include RBC membranes with randomly (and dynamically changing)
broken cytoskeletal meshworks.  As outlined in section
\ref{sec:model}, we assume the dynamics of spectrin
association and disassociation are governed by
simple rate processes.  The rate constants for
connecting a broken cytoskeletal filament, $k_c$, and
breaking an intact filament, $k_d$, are assumed independent
of the distance between filament end points on the bilayer surface
and independent of all other connections within the network.
This immediately leads to the conclusion that the probability
for a filament to be intact at a given time is $p=k_c/(k_c + k_d)$.
Equivalently, $p$ is the average percentage of intact filaments in
the cytoskeletal network.
For the moment, we simply take this picture as a hypothetical
model for non-equilibrium dynamics in the RBC membrane.
A discussion regarding the connection between this model
and experiment will be provided in sec. \ref{sec:experiments}.

Our analysis in this section centers around the calculation
of $\sigma(p)$, the effective tension of the composite membrane
as inferred from long wavelength fluctuations.  As indicated by
the notation, this tension depends on the degree of connectivity
within the network.  Not apparent in the notation is the fact that
this tension also depends upon the magnitude of the rates
$k_d$ and $k_c$ (and not just the ratio of them in $p$) due to the non-equilibrium nature of the dynamics.
Extraction of $\sigma(p)$ follows the same general prescription as in the
previous section; the fluctuation spectrum is collected and compared
to the empirical result of eq. \ref{eq:hk2}.  At the longest wavelengths
modeled in the simulations, it is found that eq. \ref{eq:hk2} does a
good job of reproducing the simulation data.  Obvious
theoretical estimates for $\sigma_{eff} = \sigma(p)$ are only available in the
limiting cases of very fast spectrin (re)association kinetics and very slow kinetics.
In general, the effective tension as a function of $p$ (and kinetic rates) must be
extracted from simulation.

In the limit that spectrin (re)association rates are much faster than all other
time scales in the problem, each cyotoskletal link in the network behaves
as an intact link with a diminished spring constant.  The numerical value
for this weakened spring constant is simply the time average of the spring as
it flips between the two possible values of $\mu$ and zero.  This picture immediately
leads to
\begin{equation}
\sigma_s(p)=p\mu,\ \ \sigma_h(p)=\sqrt{3}p\mu.  \ \ \ \ \ \ (\mathrm{fast \ \ spectrin \ \ kinetics}) \label{eq:sigmap1}
\end{equation}
Similar arguments have been invoked previously to suggest a possible ATP dependence
in the shear modulus of the RBC membrane \cite{nir_cyto}.  Such an ATP concentration
dependence may provide an explanation for RBC shape changes as a function
of ATP concentration \cite{nir_cyto}.

In the opposite regime, when network connectivity kinetics are far
slower than any other timescale in the system, the membrane may be
regarded as evolving thermally under the influence of quenched
network disorder. The behavior of the system in the quenched
disorder limit is analogous to the 2D percolation problem
\cite{Kirkpatrick}.  As such, it is expected that the effective
tension within the system must vanish at some finite critical
$p=p_c$.  For values of $p$ less than $p_c$, no global
connectivity within the network exists and, consequently, there is
no restoring force possible in response to area dilations of the
membrane surface.  The 2D connectivity percolation limit is known
to occur at $p_c=2/z$, where $z$ is the bond valence for each node
(i.e. $z=4$ with $p_c=1/2$ for the square network and $z=6$ with
$p_c=1/3$ for the triangular network). Furthermore, within the
approximation of the mean field theory, it is expected that the
decrease in $\sigma$ as  $p$ drops from one to $p_c$ is linear.
That is
\begin{eqnarray}
\sigma_s(p)=2\mu(p-1/2),\ \
\sigma_h(p)=\frac{3\sqrt{3}}{2}\mu(p-1/3)\nonumber\\  \ \ \ \ \ \
(\mathrm{slow \ \ spectrin \ \  kinetics}) \label{eq:sigmap2}
\end{eqnarray}
for all $p>p_c$ and $\sigma=0$ for $p<p_c$.
A brief justification for the percolation theory results quoted above may be found in the
Appendix.

The discussion of ``fast'' and ``slow'' kinetics in the previous paragraph was intentionally
left ambiguous, without specification of what time scales these quantities were to
be compared with.  It seems prudent to avoid a detailed discussion of this issue, due to
the many different dynamic scales in this particular problem, however a crude discussion
is appropriate.  Given our focus on long wavelength elastic properties, the most obvious
timescale for comparison is the membrane relaxation time for the longest wavelength under observation.
For equilibrium membranes at tension $\sigma$ and bending rigidity $\kappa$, it is well
known \cite{brochard,Granek} that relaxation of $h_{\mathbf{k}}$ is exponential with a characteristic time
\begin{equation}
\tau_m(k)=\frac{4\eta}{\kappa k^3+\sigma k}.
\label{eq:taum}
\end{equation}
This result follows immediately from eq. \ref{eq:langevink}.
Considering only the longest wavelength modes in our simulations
($k=2\pi/L_{max}$ with $L_{max}=L$ for the square and
$L_{max}=L_y$ for the hexagonal simulations) and allowing for
tensions between zero and the fully connected network values, we
find that $\tau_m$ falls in the range of
$9\times10^{-4}\mathrm{s}$  - $8\times10^{-3}\mathrm{s}$. While
these values are only required to hold for homogeneous equilibrium
membranes, they were verified to hold reasonably well for the
non-equilibrium membranes studied here when $\sigma(p)$
(determined from the fluctuation spectrum) is naively used in eq.
\ref{eq:taum}.  The relaxation rate associated with the two-state
spectrin dynamics is $(k_c + k_d)$, which provides a time scale
$\tau_{cd} = 1/(k_c + k_d)$.  The fast kinetics limit discussed
above would be expected to apply for $\tau_m \gg \tau_{cd}$ and
the slow kinetics limit would apply for $\tau_m \ll \tau_{cd}$.

FSBD simulations were run including KMC breaking and reforming
events in the spectrin meshwork (as detailed in sec.
\ref{sec:simulation}).  Although no equilibrium can be reached in
these simulations due to the intrinsically non-equilibrium nature
of the simulation, the systems do converge to a steady state
regime.  $\langle|h_\mathbf{k}|^2\rangle$ as a function of $k$
were collected, at steady-state, for both the square and hexagonal
systems described in the previous section.  Three different sets
of simulations were run in each geometry corresponding to
different spectrin kinetic regimes.  Specifically, $k_d$ was
chosen to assume the values  $10^4$ s$^{-1}$, $200$ s$^{-1}$, and
$10$ s$^{-1}$.  A variety of different connectivity percentages
were simulated, spanning $0<p<1$.  Given values for $k_d$ and $p$,
$k_c = k_d p/(1-p)$ follows immediately allowing for complete
specification of the model. The three choices for $k_d$ roughly
correspond to the regimes $\tau_{cd}\ll\tau_{m}$,
$\tau_{cd}\sim\tau_{m}$, and $\tau_{cd}\gg\tau_{m}$ respectively.

Figures \ref{fig:squbroke} and \ref{fig:tribroke} display our
simulation results, plotted in the form of effective surface
tensions as a function of network connectivity.  The tensions are
calculated as in the lower panel of Fig.~\ref{fig:mcmesh} using
the largest wavelengths available in the simulations.  In addition
to the FSBD/KMC simulations, FMC results are plotted for the
quenched disorder case corresponding to $(k_c + k_d)=0$.  In these
simulations, an ensemble of random network connectivities
consistent with the prescribed $p$ values were run without
allowing the network connectivity to evolve.
 The results for $\langle|h_\mathbf{k}|^2\rangle$ were averaged over the ensemble to obtain $\sigma(p)$.
 FMC was used in these cases for numerical efficiency and is valid since the evolution under conditions of
 static network connectivity is purely thermal.  The speed of the FMC algorithm allowed a set of simulations to be run
 for larger membrane sizes in order to clarify the role of finite size effects.

 The figures clearly display both expected regimes.  Fast network reorganization yields
 results in good agreement with eq. \ref{eq:sigmap1}, whereas slow reorganization yields
 results consistent with eq. \ref{eq:sigmap2}.  Intermediate kinetic regimes interpolate
 between these two limiting cases.  Finite size effects appear not to play any role, except
perhaps for the case of $p$ values very close to $p_c$ in the quenched disorder case, which
is to be expected due to the percolation phase transition at this point.  At
values of intermediate membrane connectivity, it is possible for the systems to display a
range of effective surface tension values, depending upon the relative timescales for
spectrin kinetics as compared to membrane relaxation.  The implications of this fact
will be discussed further in the next section.



\section{Connection to experiments}
\label{sec:experiments}

Evidence that RBC membrane shapes and shape fluctuations depend
upon non-thermal energy expenditure comes from two different
experimental sources.  Measurement of cell shape
\cite{Sheetz_Singer} and shape fluctuations
\cite{Korenstein, Tuvia} under a variety of MgATP
concentrations suggest this fact directly.  Increases in MgATP
concentration lead to enhanced membrane surface fluctuations.
Less directly, it has been shown that RBC membrane
fluctuation amplitudes in the presence of MgATP depend upon the viscosity of the surrounding
solvent \cite{Tuvia2}.  Thermal properties of a physical system should
not be influenced by transport coefficients, such as the viscosity, but should
depend only upon system energetics via the partition function.

Biochemically, it is clear that the presence of MgATP leads to the
phosphorylation of spectrin \cite{Birchmeier}, which has been
implicated in the observed shape changes and fluctuation
amplitudes in RBC membranes under varying MgATP concentrations
\cite{Sheetz_Singer}. Whether or not this phosphorylation event
(and/or similar events in other molecular components of the
network) coincides with breaking of spectrin filaments is less
clear, however the hypothesis that this is in fact the case
\cite{nir_cyto} is compelling and served as one of the major
motivations for this study.  In what follows, we assume that
phosphorylation of spectrin induces individual filaments to
dissociate from the network.  We stress that this model is
hypothetical and, indeed, that some studies refute the correlation
between spectrin phosphorylation and RBC shape/elastic properties
\cite{patel,moret}.

In a naive model for spectrin phosphorylation, we assume the following kinetic
equations for spectrin dissociation and association with the network as a whole.
\begin{eqnarray}
\label{eq:phosphor} \mathrm{S_a} + \mathrm{ATP} &
\stackrel{k_d}{\rightarrow} & \mathrm{S_d} + \mathrm{ADP} \\
\nonumber S_d  + \mathrm{H_2 O} & \stackrel{k_c}{\rightarrow} &
\mathrm{S_a} + \mathrm{HPO_4^{2-}}.
\end{eqnarray}
In the above, $\mathrm{S_a}$ and $\mathrm{S_d}$ stand for the
associated and dissociated forms of spectrin, respectively, and it is assumed that
the reactions proceed via catalysis by kinases and phosphatases.  We assume
ATP concentration is held fixed, either by endogenous biochemistry within the RBC
in vivo or by experimental control in vitro.
Using our previously introduced notation, we write $[\mathrm{S_a}]=Cp$
and $[\mathrm{S_d}]=C(1-p)$, where $C$ is the total concentration of spectrin on the cell surface.
The steady state condition is formulated as $d p/d t = 0$, which, in conjunction with kinetic
eqs. \ref{eq:phosphor}, leads to
\begin{equation}
\label{eq:pATP} p=\frac{n_c}{n_c+2n}.
\end{equation}
Here, both kinetic constants have been wrapped up into the single
constant $n_c$.  $n$ is the concentration of ATP in solution.
$n_c$ has been defined such that the condition $n=n_c$ implies
$p=p_c=1/3$ (in this section we assume the hexagonal network due
to its direct connection to RBC systems) , i.e. $n=n_c$ specifies
the concentration of ATP necessary to reach the percolation phase
transition in systems with quenched disorder.  While this model
neglects the observed dependence of spectrin dephosphorylation on
$n$ \cite{Birchmeier}, it has the advantage of simplicity in the
absence of quantitative kinetic data and follows the line of
reasoning previously introduced \cite{nir_cyto}, allowing for
comparison to earlier work.  Eq. \ref{eq:pATP} predicts the
connectivity of the network as a function of ATP concentration and
allows for comparison between the modeling of the previous section
and experimental data.

\begin{figure}
\begin{center}
\includegraphics[width=0.5\textwidth]{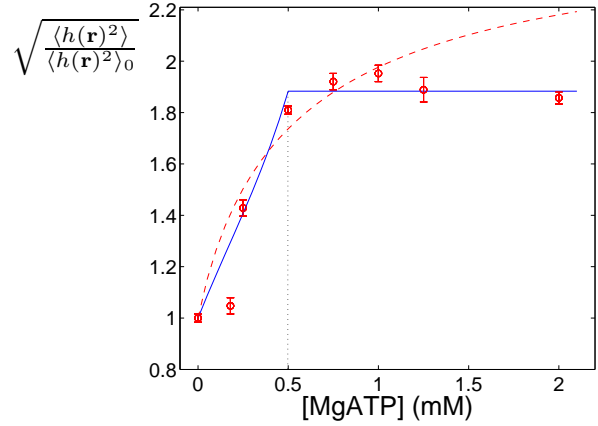}
\end{center}
\caption{Real space RBC height fluctuations, $\sqrt{\langle
h(\mathbf{r})^2\rangle}$, as a function of ATP concentration
(normalized by the zero ATP values, $\sqrt{\langle
h(\mathbf{r})^2\rangle}_0$). The circles with error bars are the
experimental data~\cite{Tuvia}. The solid line is a theoretical
fit assuming $\sigma_h$ in Eq.~(\ref{eq:sigmap2}), corresponding
to the case of $\tau_{m}\ll \tau_{cd}$ (slow spectrin kinetics).
The dashed line is a theoretical fit assuming $\sigma_h$ in
Eq.~(\ref{eq:sigmap1}), corresponding to the case of $\tau_{cd}\ll
\tau_{m}$ (fast spectrin kinetics).} \label{fig:exp}
\end{figure}

Experimental data for RBC fluctuations as a function of ATP
concentration is available only in terms of real space height amplitudes,
$h(\mathbf{r})$ \cite{Tuvia} (see fig. \ref{fig:exp}).
Following Eqs.~(\ref{eq:Fourierk}) and (\ref{eq:hk2})
(and for simplicity assuming $L_x = L_y = L$),
\begin{equation}
\langle
h(\mathbf{r})^2\rangle=\sum_{\mathbf{k}}\frac{k_BT}{L^2(\kappa
k^4+\sigma k^2)},\label{eq:hr}
\end{equation}
where the interval between adjacent $k_x$ or $k_y$ is $2\pi/L$. In
the limit of large $L$, the sum $\sum_{\mathbf{k}}$ can be
approximated by the integral $L^2/(2\pi)^2\int d\mathbf{k}$, and we
have
\begin{equation}
\langle
h(\mathbf{r})^2\rangle=\frac{k_BT}{4\pi\sigma}\ln(1+\frac{\sigma
L^2}{\pi^2\kappa}).\label{eq:realh}
\end{equation}

The limiting expressions and simulations of the preceding section
provide a route toward estimation of $\sigma(p)$ under various
rates of spectrin association/dissociation.  Eq. \ref{eq:pATP}
enables us to translate these results into tensions as a function
of ATP concentration. The resulting tensions may be used in eq.
\ref{eq:realh} to calculate the real space membrane fluctuation
amplitudes  as a function of ATP concentration.  RBC's have a
diameter of 7 microns and we use $L=7 \mathrm{\mu m}$~\cite{lodish}
in our comparison to experiment.  It is also important to note
that experimental observations of RBC fluctuations were carried out
on cells that were allowed to
 ``firmly adhere to [a] glass substratum" \cite{Tuvia}
under the influence of their natural (presumably of electrostatic
origin) affinity for such surfaces.  The adhesion of the bilayer
to an underlying supporting matrix will be assumed to contribute a
bare surface tension, $\sigma_{bare}$ to the membrane of the
RBC \cite{reservoir,brown_groves}.  That is, we assume
$\sigma = \sigma_{bare} + \sigma(p)$.

We take $n_c$ and $\sigma_{bare}$ as two fitting
parameters to reproduce the experimental data.  The resulting
fits, assuming the two extreme cases of fast spectrin kinetics
(eq. \ref{eq:sigmap1} assumed) and slow spectrin kinetics (eq.
\ref{eq:sigmap2} assumed) are displayed in fig. \ref{fig:exp}. The
values adopted by the fitting parameters in these two cases are
$\sigma_{bare}=0.2\mu=150 \mathrm{k_BT/\mu m^2}$ (both cases), and $n_c=0.22$
mM for the fast kinetics case and $n_c=0.5$ mM for the slow
kinetics case. It turns out that the limiting case of slow
spectrin kinetics provides the most satisfactory fit of the data,
not only in comparison to the fast kinetics case, but also with
respect to intermediate kinetic regimes; the sharp onset of
tensionless fluctuations at $n_c$ naturally leads to the observed
plateau in the experimental data.


The  slowest possible time-scale associated with RBC membrane fluctuations
in our model ($\tau_m$) is on the order of a couple of seconds.  This assumes a tension-free
membrane and $L=7\mu$m.  Our data analysis suggests that spectrin breaking/reformation
dynamics are substantially slower than this, since the quenched-disorder (percolation
theory) results provide the best fit to the experimental data.  This finding would appear
to be consistent with the experimental observation  \cite{Birchmeier} that establishment
of steady-state levels of spectrin phosphorylation occurs on the time scale of minutes to
tens of minutes with changes in MgATP concentration.  Previous theoretical work
exploring the consequences of spectrin dissociation on membrane fluctuations \cite{nir_cyto}
has assumed the time-scale for spectrin kinetics to be on the order of milliseconds.  Within
the context of the present model, rates this fast seem unlikely.

Experimentally, it is observed that RBC fluctuation amplitudes
depend not only upon MgATP concentration, but also upon the
viscosity of the solvent surrounding the membrane~\cite{Tuvia2};
increasing viscosity above that of aqueous buffer solution leads
to decreases in observed fluctuations.
Membrane fluctuations at thermal equilibrium should yield
identical statistics regardless of solvent viscosity (although the dynamics
will certainly differ), so these experiments provide additional proof
of the non-thermal character of RBC fluctuations.

Within the
present model, Eq.~(\ref{eq:taum}) clearly displays the linear
relationship between inverse viscosity and the relaxation timescales for
individual membrane modes.  In the preceding section, it was
argued that the ratio between membrane relaxation rates and
spectrin kinetic rates determines the magnitude of effective tension
that must be used if one desires to fit membrane fluctuations to
the functional form of eq. \ref{eq:hk2}.  At a given connectivity,
$p$, fluctuation amplitudes decrease ($\sigma_{eff}$ increases)
as the spectrin kinetic cycle increases in speed.  Alternatively,
since it is the ratio of kinetics to membrane relaxation that is relevant,
we expect fluctuation amplitudes to decrease as viscosity is raised
assuming all other system parameters are held fixed.  Qualitatively,
this trend is in agreement with the experimental results.

To obtain quantitative comparison with the variable viscosity data
it is necessary to evaluate $\langle h(\mathbf{r})^2\rangle$ explicitly
from direct simulation, without assuming a single effective tension
as in eq. \ref{eq:hr}.  Within our model, fluctuation amplitudes are
affected by viscosity changes only by shifting the timescale for
membrane dynamics relative to spectrin kinetics.  A relative shift
in timescale (at constant $p$) corresponds to a vertical motion between the limiting
regimes plotted in fig. \ref{fig:tribroke}.  In order for viscosity changes
to yield any effect, the timescales for spectrin kinetics and membrane
relaxation must be comparable.  (e.g. if spectrin dissociation takes an hour,
increasing viscosity by a factor of 10 will not remove you from the limiting
``slow kinetics regime''.  Similarly, if dissociation took a nanosecond, no
reasonable viscosity change could promote the system out of the ``fast kinetics regime''.)
And, if these timescales are comparable, differences in the
individual relaxation rates at each wavelength will lead to different effective
tensions as a function of wavelength.  In practice, it is not feasible to run
simulations for a full $7 \mu m\times 7\mu m$ patch with sufficient sampling
to obtain reliable results.  It is possible to run a $1 \mu m\times 7\mu m$ patch.
The effective tensions as a function of wavevector were extracted from this
rectangular geometry and were used in a generalized version of eq. \ref{eq:hr}
with $\sigma \rightarrow \sigma(k)$ to obtain $\langle h(\mathbf{r})^2\rangle$ for the full
$7 \mu m\times 7\mu m$ system.

\begin{figure}
\begin{center}
\includegraphics[width=0.5\textwidth]{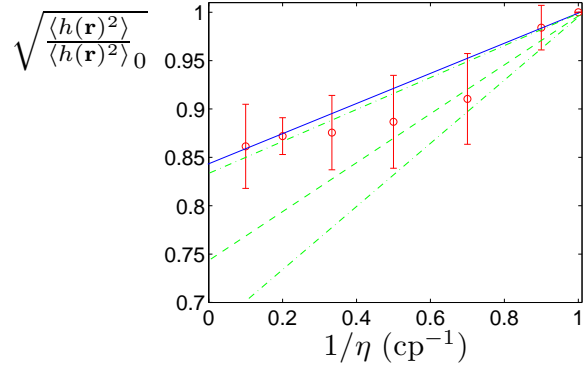}
\end{center}
\caption{Normalized height fluctuations, $\sqrt{\langle
h(\mathbf{r})^2\rangle}$, as a function of inverse viscosity of
the solvent. The circles with error bars are the simulation
results assuming $L= 7$ $\mu$m, $\sigma_{bare}=150
\mathrm{k_BT/\mu m^2}$, $n_c=0.5$ mM, $n=1.2$ mM and
$\tau_{cd}=5\times 10^5\mu$s. The solid line is a fit to our
simulation data. The dashed line and the two dashdotted lines are
the experimental fitting line and the range of experimental errors
taken from Fig. 2 of Ref.~\cite{Tuvia2}.}\label{fig:exp2}
\end{figure}

To simultaneously find agreement with the ATP concentration data (fig. \ref{fig:exp})
and the variable viscosity data it is necessary to assume spectrin rates that are
slow relative to membrane relaxation when the solvent is pure buffer solution (as
in the conditions of ref. \cite{Tuvia} and fig. \ref{fig:exp}),
but are poised to move out of this limiting regime with small viscosity increases.
By trial and error, it was found that $\tau_{cd}=5\times 10^5\mu$s for
$n=1.2$ mM (the ATP concentration conditions of ref. \cite{Tuvia2}) satisfies
this condition and yields the best fit to both data sets.  Other model
parameters were taken identical to the ``slow spectrin kinetics" fit
to the data of fig. \ref{fig:exp}.  In particular, $L= 7\mathrm{\mu m}$, $n_c=0.5$ mM and
$\sigma_{bare}=150 \mathrm{k_BT/\mu m^2}$.  In fig.~\ref{fig:exp2}
we plot real space membrane fluctuations as a function of $\eta^{-1}$ for
viscosity values spanning an order of magnitude.  We observe a
reduction in fluctuation amplitudes of about 20\% over the entire range of
viscosities studied, which falls just outside the errors of the
experimental data \cite{Tuvia2}.  It is important to stress that
figs. \ref{fig:exp} and \ref{fig:exp2}  simultaneously fit two very different experiments
to a single set of physical parameters.  Our results agree with both experiments
quite well.


\section{Discussion}\label{sec:conclusion}

From a mathematical standpoint, the results of the preceding section
could be viewed as a success; we fit the available experimental data with
our theoretical model.  However, the physical implications of
the derived fit parameters are troubling.  Most worrisome is the critical
concentration of ATP for onset of the percolation limit, $n_c=0.5$ mM.  This
implies that at physiological conditions ($n\sim 2$ mM), the spectrin network
is 89\% dissociated.  A similar degree of dissociation (83 \%) is predicted for the
$n=1.2$ mM conditions of the variable viscosity experiments described above.
As presented, it is clear that our model should not be taken seriously in the limit of slow spectrin
kinetics and $n \geq n_c$, since we do not allow the connectivity of the network to evolve
away from its starting point (i.e. a given spectrin tetramer is assumed to always link
the same two vertices or be disassociated).  In a severely compromised network, how
could a given filament ever be expected to find its assigned association point following
a dissociation event?

As mentioned previously, there does not appear to be definitive experimental data relating ATP
consumption to spectrin network dissociation.  Rather, this is a plausible hypothesis advocated
in reference \cite{nir_cyto}, based upon the limited biochemical data that is available for the various
constituents of the RBC cytoskeleton.  We can not reconcile this picture with the simulations carried
out in this work, primarily for the reason outlined in the previous paragraph.  It is possible
that ATP consumption could lead to a significant change in the elastic properties of individual
spectrin filaments without completely dissociating the filament from the network.  In such a
picture, the two forms of spectrin introduced above, $S_d$ and $S_a$, would correspond
to filaments with weak and
strong effective spring constants, respectively (or different natural lengths, etc.).
In this context, the analysis we present is sensible since network connectivity is always maintained,
but the elastic properties of individual links are changing.  The percolation limit in this picture corresponds to the point at which it is impossible
to follow a path across the spectrin network without stepping on at least one weak $S_d$ link.
Although such a picture is completely hypothetical, we note that it is theoretically possible
to alter the effective elastic properties of spectrin filaments while maintaining lateral integrity of
the network \cite{nir_njp,asaro}.

The underlying physical picture pursued in this work was motivated by and is similar in spirit
to the work of Safran and Gov \cite{nir_cyto} (i.e. a dynamically breaking and reforming
spectrin network).  However,  the simulation model implemented to study this system
is more similar to the work of Dubus and Fournier \cite{Fournier} (with the additional possibility of dynamically breaking bonds).  We have interpreted our simulation results within the context of a simple percolation theory.  While this picture shares some similarity with the analysis of ref. \cite{nir_cyto}
in the context of global effects associated with variable ATP concentration, we find differences
between the present model and ref. \cite{nir_cyto} in the description of local fluctuation
dynamics.  For example, we find
no evidence in our simulations for localized forces of the sort discussed in ref. \cite{nir_cyto}.  Instead,
the increased amplitudes of fluctuation at high ATP concentration are attributable to global properties of the spectrin network within our model.  We also find that the rates
associated with spectrin kinetics must be orders of magnitude slower than assumed in ref. \cite{nir_cyto}
in order to reconcile our simulation model with the experiments of refs. \cite{Tuvia,Tuvia2}.  However,
it should be pointed out that the particular rates assumed in ref. \cite{nir_cyto} were only
rough estimates and are not essential to the qualitative predictions of that work \cite{nir_talk}.

The starting point for all of our simulations, eq. \ref{eq:H},
assumes a central-force network for the cytoskeletal matrix and a
phantom entropic spring description for the spectrin filaments.
In addition, we do not allow for the possible formation of 5-fold
or 7-fold defects \cite{nelson_defect,nir_cyto} in our network.
Inclusion of direct interactions between spectrin and the bilayer
surface, generalizing the description of the network to deviate
from the central-force picture, and/or allowing for defects in
network connectivity could all potentially alter the quantitative
findings of this work.  The present simulation model was adopted
for ease of numerical implementation and to facilitate comparison
to the existing work of Dubus and Fournier \cite{Fournier}.  This
study provides a detailed analysis of the behavior of a specific
model for a composite membrane system (lipid bilayer plus actively
labile cytoskeletal network).  This model is motivated by our
(incomplete) picture for the structure and kinetics of the RBC
membrane surface and may prove useful in the development of more
refined models in the future.  In particular, it should be noted that more
refined descriptions of the spectrin network have been developed that
allow for interactions between the bilayer surface and the filaments and
a more complete description of polymer elasticity beyond the simple
Gaussian-chain model \cite{boal94,suresh,sung}.  Future modeling,
incorporating some of these approaches for the behavior of the cytoskeleton,
would be highly desirable and is currently under investigation.

We note in closing that a very recent study by Evans et. al. \cite{kimparker} directly
contradicts the experimental results in refs. \cite{Tuvia,Tuvia2}.  This recent series of
experiments finds no correlation between ATP concentration and RBC membrane fluctuations.
Given the severe disagreement between experimental results on the RBC system, one has
almost complete freedom in interpretation of our simulation results.  The picture outlined
above suggests that we can obtain quite good agreement between simulation and experiments
that do measure ATP dependence of the RBC fluctuations.  Within this picture, it seems
necessary to assume that our ``broken" spectrin links actually correspond to intact links
with diminished spring constant and/or a non-zero natural length.  Another, equally valid,
interpretation of our results is that it is impossible to reconcile an actively breaking
cytoskeletal meshwork picture with available experimental data due to the high degree of network dissociation predicted under physiological conditions within such a picture.
Given the uncertain experimental landscape, it seems impossible to make a definitive statement
in favor of either viewpoint.

\begin{acknowledgments}
We thank Golan Bel, Nir Gov and Kim Parker for useful discussions.
This work was supported by the National
Science Foundation (grant No. CHE-0321368 and grant No. CHE-0349196).
F. B. is an Alfred P. Sloan research fellow and a Camille Dreyfus Teacher-Scholar.
\end{acknowledgments}

\section*{Appendix}
\label{sec:percolation}

The calculation of an effective tension in the limit of quenched
disorder within the spectrin network is closely related to the 2D
bond percolation problem~\cite{Stauffer}.  A mean field treatment
for the percolation problem was first developed in the context of
electric conductivity~\cite{Kirkpatrick} and was later extended to
calculate the elastic moduli of networks of Hookean springs with
finite~\cite{Feng,Feng2} and zero natural
lengths~\cite{Thorpe1,Thorpe2}. The below discussion summarizes
the arguments of Ref.~\cite{Feng} for the readers' convenience.

\begin{figure}
\begin{center}
\includegraphics[width=0.5 \textwidth]{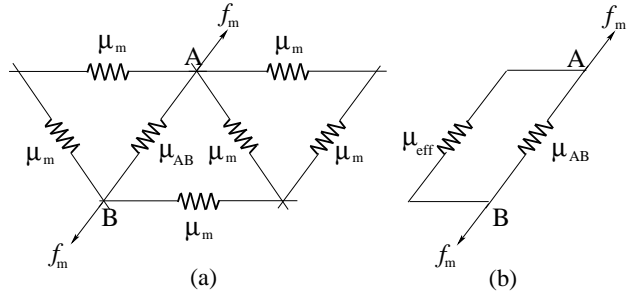}
\end{center}
\caption{(a) A randomly broken network of springs is approximated
by a complete meshwork of springs with diminished spring constant
$\mu_m$. Here we imagine an attempt to extend the link $AB$ using
a force of $f_m$. (b) The force $f_m$ is resisted both directly by
link $AB$ and the rest of the meshwork. The latter can be
considered as a single spring with effective spring constant
$\mu_{eff}$, which acts in parallel with $AB$.  (Adapted from ref.
\cite{Feng})} \label{fig:meshforce}
\end{figure}

Let us consider a randomly connected meshwork with only a fraction ($p$) of the links
intact.  Each intact link is a spring with spring constant
$\mu$ and a natural extension of zero.
In the spirit of mean field theory, we assume that the global elastic
properties of this imperfect meshwork are well approximated by
a complete meshwork comprised of springs with spring constant $\mu_m$ at every link
(see Fig.~\ref{fig:meshforce}) even though individual links will be either completely intact or
fully broken.  Our problem is to determine an expression for $\mu_m$.
For this purpose, let us consider an arbitrary link $AB$ in the
meshwork. The real spring constant, $\mu_{AB}$, associated with
this link is either $\mu$ if
the link is connected, or 0 if it is not. If we apply a force
$f_m$ across this link (see Fig.~\ref{fig:meshforce}),
the extension $x_{AB}$ is given by
\begin{equation}
x_{AB}=\frac{f_m}{\mu_{AB}+\mu_{eff}},
\end{equation}
where $\mu_{eff}$ is an effective spring constant reflecting the
contribution of all network components except the link $AB$. It
can be shown that~\cite{Feng}
\begin{equation}
\mu_{eff}=(z/2-1)\mu_m,\label{eq:mueff}
\end{equation}
where $z$ is the connectivity of the meshwork, for example, $z=4$
for the square meshwork and $z=6$ for the hexagonal meshwork
considered in this work.

On the other hand, if we assume the mean spring constant $\mu_m$ for
link $AB$, we predict a mean extension,
\begin{equation}
x_m=\frac{f_m}{\mu_m+\mu_{eff}}.
\end{equation}
To achieve consistency within the mean field approximation,
we must have
\begin{eqnarray}
\label{eq:avax}
\langle x_{AB}\rangle &=& x_m \\ \nonumber
\frac{p f_m}{\mu + \mu_{eff}} + \frac{(1-p)f_m}{\mu_{eff}} &=& \frac{f_m}{\mu_m + \mu_{eff}}
\end{eqnarray}
where the second equality is simply a more explicit
version of the first.  This equation is readily solved
to yield
\begin{equation}
\mu_m=\frac{p-p_c}{1-p_c}\mu,\label{eq:mum}
\end{equation}
where
\begin{equation}
p_c=2/z.\label{eq:pc}
\end{equation}
To calculate $\sigma(p)$, we simply replace $\mu$ in
eq.~\ref{eq:sigma} by $\mu_m$ from eq.~\ref{eq:mum}, since the
latter is now the average spring constant of the equivalent
complete meshwork. Taking $z=4$ and $z=6$ for the square and
hexagonal meshwork respectively, we arrive at
eq.~\ref{eq:sigmap2}.

\begin{figure}
\begin{center}
\includegraphics[width=0.5\textwidth]{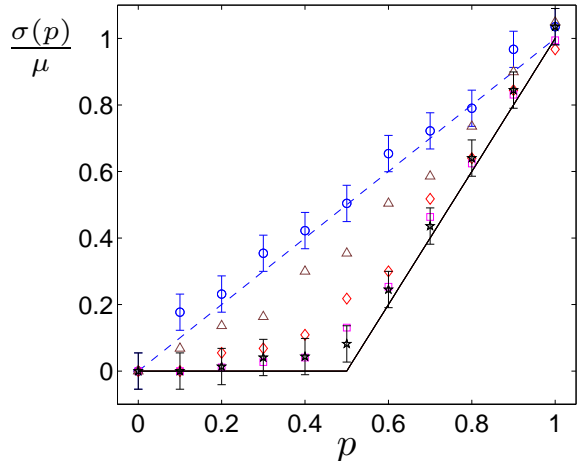}
\end{center}
\caption{$\sigma(p)/\sigma(p=1)$ versus $p$ is plotted for a
square meshwork. The circles, triangles, and diamonds are FSBD/KMC
results for $k_d=10^4 s^{-1}$ ($\tau_{cd}\ll\tau_{m}$), $k_d=200
s^{-1}$ ($\tau_{cd}\sim\tau_{m}$) and $k_d=10 s^{-1}$
($\tau_{cd}\gg\tau_{m}$) respectively. The squares and the stars
are FMC results ($k_d = k_c=0$).  All simulations use box sizes
identical to those introduced in the previous section, except for
the stars.  Stars represent FMC simulations on systems that are
twice as large in both linear dimensions. For clarity, only error
bars for the circles and the stars are shown, but all simulations
have similar errors. The dashed line is the prediction of
Eq.~(\ref{eq:sigmap1}). The solid line is the prediction of
Eq.~(\ref{eq:sigmap2}).}\label{fig:squbroke}
\end{figure}

\begin{figure}
\begin{center}
\includegraphics[width=0.5\textwidth]{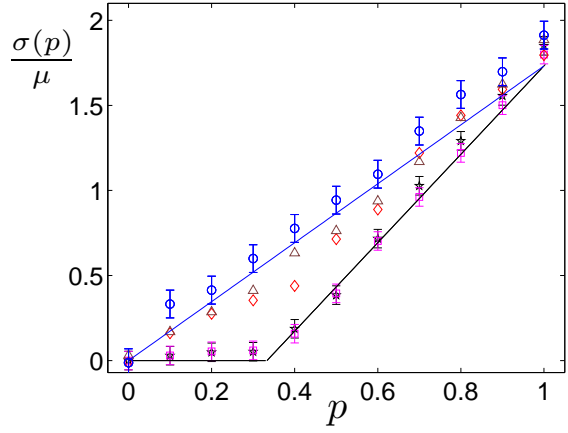}
\end{center}
\caption{$\sigma(p)/\sigma(p=1)$ versus $p$ is plotted for a
hexagonal meshwork. All symbols and lines have the same meaning as
in Fig.~\ref{fig:squbroke}, only the meshwork connectivity has
been changed.  (The obvious discrepancy between simulation and theory
at $p=1$ is due to the rectangular geometry of our simulation box.  It is impossible
to perfectly fit a portion of a hexagonal network within a rectangle, so the
theoretical results that assume perfect hexagonal symmetry are only approximately
valid in this case.)}\label{fig:tribroke}
\end{figure}

One should note that eqs.~\ref{eq:sigmap2} and \ref{eq:pc} are
only results of a mean field approximation. In the vicinity of
$p_c$, it is well known that the correlation effects are strong
and that mean field theory breaks down. However, the mean field
results are adequate to describe the behavior of $\sigma$ outside
the immediate vicinity of $p_c$ as evidenced by figs.
\ref{fig:squbroke} and \ref{fig:tribroke}, which is sufficient for
the purposes of this work.

\newpage
\begin{table*}[h]
\caption{Physical parameters in simulations}
\begin{tabular}{cccc}
\hline
Parameter \ \ & Description \ \ & Value \ \ & Reference \\
$\kappa$ \ \ & bending modulus \ \ & 4.8
k$_\mathrm{B}$T\footnotemark
 \ \ & \cite{brochard}   \\
$\eta$ \ \ & cytoplasm viscosity \ \ & 1.5 k$_\mathrm{B}$T s
$\mu$m$^{-3}$
\ \ & \cite{brochard} \\
$D$ \ \ & diffusion constant of nodes \ \ & 0.53 $\mu$m$^2$
s$^{-1}$\ \
& \cite{Tomishige}\\
$A$ \ \ & the average distance between two nearest nodes (see
Fig.~\ref{fig:corral}) \ \ &
0.112 $\mu$m \ \ & \cite{Boal} \\
$\ell_c$ & contour length of the spectrin tetramer \ \ & 0.2
$\mu$m &
\cite{Boal}\\
$\ell_k$ & Kuhn length of the spectrin tetramer \ \ & 0.02 $\mu$m
&
\cite{Boal}\\
$\mu$ \ \ & spring constant of the spectrin tetramer\ \ & 750
k$_\mathrm{B}$T
$\mu$m$^{-2}$ & \footnotemark \\
\hline
\end{tabular}
\label{tab:para} \footnotetext[1]{T$\equiv$ 300K.}
\footnotetext[2]{Calculated using the Gaussian chain model through
two parameters above.}
\end{table*}

\newpage

\newpage

\newpage

\newpage

\newpage

\newpage

\newpage

\newpage

\newpage

\newpage

\newpage

\newpage

\end{document}